\UseRawInputEncoding
\documentclass[twocolumn,amsmath,aps]{revtex4}

\usepackage{amssymb,mathrsfs,bm,color,times}
\usepackage{graphicx,color}
\usepackage{bm}
\usepackage[hypertex]{hyperref}
\usepackage{amsmath}

\newcommand{\be}{\begin{equation}}
\newcommand{\ee}{\end{equation}}
\newcommand{\bea}{\begin{eqnarray}}
\newcommand{\eea}{\end{eqnarray}}
\newcommand{\bsube}{\begin{subequations}}
\newcommand{\esube}{\end{subequations}}

\newcommand{\Eq}[1]{Eq.\,(\ref{#1})}

\newcommand{\la}{\langle}
\newcommand{\ra}{\rangle}

\newcommand{\beq}{\begin{equation}}
\newcommand{\eeq}{\end{equation}}
\newcommand{\beqn}{\begin{eqnarray}}
\newcommand{\eeqn}{\end{eqnarray}}

\newcommand{\nl}{\nonumber \\}

\newcommand{\bsub}{\begin{subequations}}
\newcommand{\esub}{\end{subequations}}

\begin{document}


\title{Dynamics simulation of braiding two Majorana-zero-modes via a quantum dot}

\author{Luting Xu}
\email{xuluting@tju.edu.cn}
\affiliation{Center for Joint Quantum Studies and Department of Physics,
School of Science, \\ Tianjin University, Tianjin 300072, China}

\author{Jing Bai}
\affiliation{Center for Joint Quantum Studies and Department of Physics,
School of Science, \\ Tianjin University, Tianjin 300072, China}

\author{Wei Feng}
\affiliation{Center for Joint Quantum Studies and Department of Physics,
School of Science, \\ Tianjin University, Tianjin 300072, China}

\author{Xin-Qi Li}
\email{xinqi.li@tju.edu.cn}
\affiliation{Center for Joint Quantum Studies and Department of Physics,
School of Science, \\ Tianjin University, Tianjin 300072, China}

\date{\today}

\begin{abstract}
In this work we perform real time simulations for the dynamics
of braiding a pair of Majorana zero modes (MZMs)
through a quantum dot in a minimal setup of pure 1D realization.
We reveal the strong nonadiabatic effect
when the dot energy level approaches to zero
in order to achieve a geometric phase $\pi/4$
which is required for a full exchange between the MZMs.
Rather than the strategies of nonuniformly manipulating the system
according to adiabatic condition and shortcuts-to-adiabaticity,
we propose and illustrate a more feasible scheme
to suppress the nonadiabatic transition,
meanwhile which allows for a full exchange between the Majorana modes.
\end{abstract}


\maketitle

{\flushleft\it Introduction}. --- The nonlocal nature of
the Majorana zero modes (MZMs) and non-Abelian exchange statistics obeyed
provide an elegant paradigm of topological quantum computation (TQC)
\cite{Kit01,Kit03,Sar08,Ter15,Sar15,Opp20}.
The exchange (braiding) operations in real space
can lead to unitary rotations in the degenerate subspace of ground states,
which constitute desired quantum information processing and realize logic gates in TQC.
In the past decade, after great efforts, considerable progress
has been achieved for realizing the MZMs in various experimental platforms.
Yet, the main experimental evidences are associated with
the zero-bias conductance peaks,
which cannot ultimately confirm the realization of MZMs.
Further step of demonstrating the existence of MZMs
and the key procedure towards TQC
is illustrating the novel non-Abelian statistics.

Viewing that the hybrid semiconductor-superconductor devices,
e.g., the $s$-wave superconductor proximitized nanowires,
have been the leading platform
to generate the MZMs\cite{Kou12,MarA16,MarD16},
thus an obvious difficulty is that directly exchanging (braiding)
the MZMs realized in 1D nanowires is impossible,
since collisions between the MZMs during braiding cannot be avoided in 1D case.
Schemes to combine the 1D wires into a 2D network (through T- or Y-junctions)
for gate-voltage-controlled moving and exchanging the MZMs
have been proposed \cite{Fish11,Opp12,Opp13,Plug17,Roy19,Han20,Li22},
yet progress to resolve the huge technological challenges (necessary controls) is slow.
Other proposals of braiding the MZMs include tuning local couplings
between Majorana modes directly via electric gates
\cite{Tew11a,Tew11,Sato15,Sau16}
or indirectly via modulating the role of charging energy
on the Majorana islands \cite{Bee12,Bee13,Beri22,Nay16}
or through quantum dots \cite{Flen11,Mora18,Flen22},
measurement-only schemes \cite{Nay08,Los16,Fre17,Fu17}, and others \cite{Naz21}.

\begin{figure}
  \centering
  \includegraphics[scale=0.6]{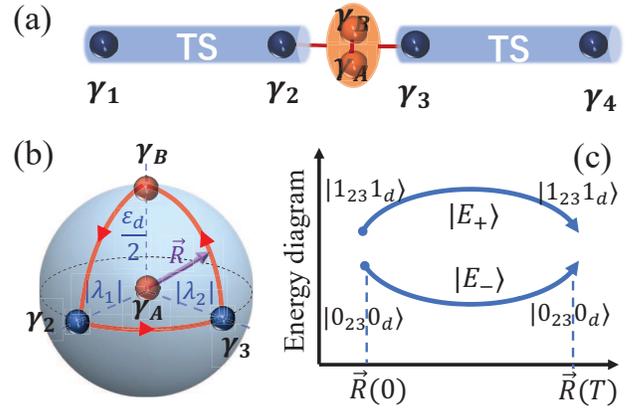}
  \caption{(a)
Schematic diagram for the setup of topological superconductor (TS) nanowires
connected with a quantum dot, proposed to braid a pair of MZMs in 1D case.
(b)
By splitting the dot electron into a pair of
Majorana fermions ($\gamma_A$ and $\gamma_B$),
the MZMs $\gamma_2$ and $\gamma_3$
and the quantum dot can be mapped to a Y-junction,
which supports formation of a geometric Berry phase,
for the use of braiding (exchanging) the Majorana modes $\gamma_2$ and $\gamma_3$.
(c)
Energy diagram of the instantaneous eigenstates $|E_{\pm}\ra$
in the subspace of even fermion parity.
Here, the occupation state $|n_{23} n_d\ra$ is used,
with $n_{23}$ and $n_d$ the occupation numbers of
the regular fermions associated with the MZMs $\gamma_2$ and $\gamma_3$
and the quantum dot, respectively.  }
\end{figure}

Very recently, following the braiding protocol via superconducting Y-junctions \cite{Fra17,Yang19,Sau19,Sar17,Mar22,Fre16} ,
an alternative scheme based on more conventional elements,
i.e., superconductor proximitized nanowires connected with a quantum dot (QD),
was proposed to braid a pair of MZMs in 1D case,
as schematically shown in Fig.\ 1(a).
In the original Y-junction proposal,
three Cooper pair boxes connected at a Y-junction
via three overlapping Majorana fermions
(which effectively produce a single zero-mode at the center).
This was regarded as the minimal setup
required for the braiding of a pair of MZMs,
controlled by the fluxes through the three Josephson junctions to a bulk superconductor.
In Ref.\ \cite{Xie21}, the setup is simplified to
a single 1D Josephson junction, as shown in Fig.\ 1(a),
where two proximitized nanowires are connected through a quantum dot.
By tuning the coupling strengths between the MZMs and the QD
through gate voltages, the MZMs $\gamma_2$ and $\gamma_3$
can be exchanged with the help of the QD.
In this work we perform real time simulations
for the braiding dynamics of the MZMs in this desired 1D setup.
In particular, we will reveal the strong nonadiabatic effect
when the dot energy level approaches to zero
in order to achieve a geometric phase $\pi/4$,
as required for a full exchange between the MZMs.
We will also simulate and compare a variety of manipulating schemes,
to exploit proper strategy to suppress the nonadiabatic transition
and achieve full exchange between the Majorana modes.

\vspace{0.1cm}
{\flushleft\it Basic idea and problem}. ---
Let us consider the setup schematically shown in Fig.\ 1(a),
where each topological superconductor wire can be realized
by a semiconductor nanowire
in proximity contact with an $s$-wave superconductor.
For each wire, a pair of MZMs appear at the ends.
Here we denote the four MZMs as $\gamma_1$, $\gamma_2$, $\gamma_3$ and $\gamma_4$.
Actually, using four MZMs is the minimal construction of a Majorana logic qubit,
in order to conserve the fermion parity (even or odd).
For the setup in Fig.\ 1(a), if combining ($\gamma_1,\gamma_2$)
as a regular fermion with occupation $n_{12}=0$ or 1,
and ($\gamma_3,\gamma_4$)
as another regular fermion with occupation $n_{34}=0$ or 1,
the 4-MZMs qubit (with even fermion parity, for example)
has logic states $|0_{12}0_{34}\ra$ and $|1_{12}1_{34}\ra$.
In Fig.\ 1(a), the QD inserted between the wires is used to mediate
exchange between the MZMs $\gamma_2$ and $\gamma_3$.
If a full exchange is accomplished,
i.e., $\gamma_2\to \gamma_3$ and $\gamma_3\to -\gamma_2$
associated with a braiding operator $U=e^{\frac{\pi}{4}\gamma_2\gamma_3}$,
the Majorana qubit state would experience a rotation as
\bea\label{psi-g}
|0_{12}0_{34}\rangle \to
|\psi_g\ra = \frac{1}{\sqrt{2}}(|0_{12}0_{34}\rangle-i|1_{12}1_{34}\rangle)  \,,
\eea
which corresponds to a Hadamard-type logic gate
(up to an additional phase shift $\pi/2$).

The QD is assumed to have only a single level
involved in the braiding dynamics,
with thus a single level Hamiltonian
$H_D=\varepsilon_d d^{\dagger} d$.
The QD is coupled to the MZMs $\gamma_2$ and $\gamma_3$,
described as
$ H'_1= i(\lambda_1 d+\lambda_1^*d^\dagger)\gamma_2$
and $H'_2=(\lambda_2d-\lambda_2^*d^\dagger)\gamma_3$.
The coupling amplitudes are more explicitly specified as
$\lambda_1=|\lambda_1|e^{i\phi_1/2}$
and $\lambda_2=|\lambda_2|e^{i\phi_2/2}$,
with the phases $\phi_1$ and $\phi_2$ modulated
by controlling the phases of the $s$-wave superconductors
through magnetic fluxes (based on the Aharonov-Bohm effect)
as explored in Refs.\ \cite{Ren17,RenN19,RenP19}.
If we decompose the dot electron into a pair of Majorana fermions
$\gamma_A$ and $\gamma_B$, described as $d=(\gamma_B-i\gamma_A)/2$,
and modulate the phases as $\phi_1=\pi$ and $\phi_2=0$,
we can easily check that
$\gamma_B$ is decoupled with the MZMs $\gamma_2$ and $\gamma_3$,
and the remained coupling can be reexpressed as
\bea
H_M = i\gamma_A(\vec{R}\cdot\vec{\gamma}) \,.
\eea
Here $\vec{R}=(|\lambda_1|,|\lambda_2|,\varepsilon_d/2)$
and $\vec{\gamma}=(\gamma_2,-\gamma_3,\gamma_B)$
are introduced for the sake of brevity.
Satisfactorily, in this way, we have mapped the setup to
the configuration of a Y-junction
\cite{Fra17,Yang19,Sau19,Sar17,Mar22,Fre16}, as shown in Fig.\ 1(b).
Through control of the coupling strengths $|\lambda_1|$ and $|\lambda_2|$
the system can complete a cyclic evolution in the parameter space,
which is described by $U=e^{\Omega\gamma_2\gamma_3}$.
Here $\Omega$ is the well-known Berry phase,
which is half of the solid angle ($\pi/2$, in the case $\varepsilon_d\to 0$)
enclosed by the evolution trajectory in the parameter space.
The result $\Omega=\pi/4$ corresponds to
a full exchange between $\gamma_2$ and $\gamma_3$,
and as well the state rotation of the 4-MZMs qubit as mentioned above.

From Fig.\ 1(b) we know that,
in order to precisely realize the solid angle $\pi/2$,
we should make $\varepsilon_d \to 0$.
However, this will cause strong nonadiabatic transitions
during the initial and final stages of the cyclic evolution,
if the speed of evolution is not slow enough.
The reason can be understood from the energy diagram depicted in Fig.\ 1(c),
for the instantaneous eigenstates.
Here, the occupation state $|n_{23} n_d\ra$ is used,
with $n_{23}$ and $n_d$ the occupation numbers of
the regular fermions associated with the MZMs $\gamma_2$ and $\gamma_3$
and the quantum dot, respectively.
Without loss of generality, we only consider
the subspace of even fermion parity,
i.e., the subspace expanded by the basis states
$|0_{23} 0_d\ra$ and $|1_{23} 1_d\ra$.
We see that, near the beginning and end of the braiding operation,
the energy difference of the instantaneous eigenstates $|E_+\ra$ and $|E_-\ra$ is $\varepsilon_d$,
noting that there is no direct coupling between $\gamma_2$ and $\gamma_3$.
This near-zero energy gap (when $\varepsilon_d\to 0$)
will make the two-state system suffer strong nonadiabatic transition,
unless the evolution is infinitely slow.
In this work we would like to investigate
the nonadiabatic effect caused by the small $\varepsilon_d$.
In particular, we will explore strategies to suppress the nonadiabatic transitions
and compare the results of different schemes,
including four-step uniform and nonuniform manipulations of the coupling parameters,
and an improved six-step manipulating scheme.

\begin{figure}
  \centering
  \includegraphics[scale=0.8]{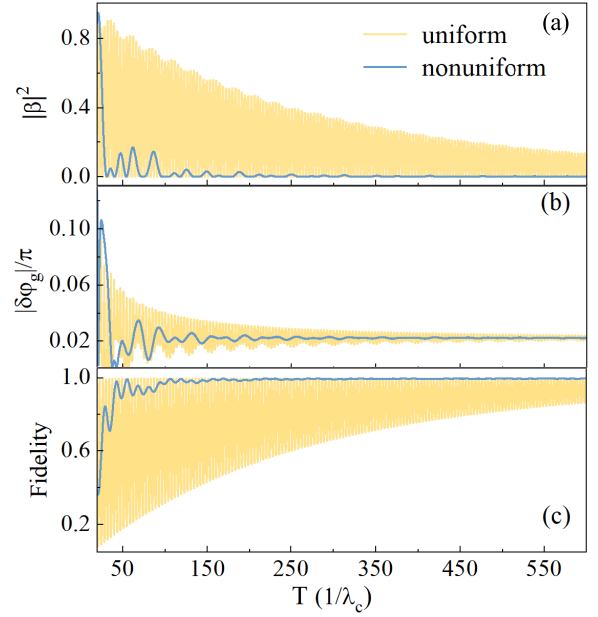}
  \caption{
Nonadiabatic effect of the four-step uniform (yellow lines) and nonuniform (blue lines)
manipulation schemes, shown by the total braiding time dependence
in (a) for the transition probability,
in (b) for the geometric phase error from the Berry phase $\pi/4$,
and in (c) for the fidelity of the 4-MZMs qubit state rotation.
In all simulations of this work, arbitrary unit of energy is assumed
by setting the maximum coupling strength $\lambda_c=1.0$.
In this plot, the quantum dot energy level is assumed as
$\varepsilon_d=0.2\lambda_c$.  }
\end{figure}

\vspace{0.1cm}
{\flushleft\it Four-step uniform and nonuniform schemes}. ---
Let us start with the four-step uniform manipulating scheme.
The basic idea of braiding is clear as shown in Fig.\ 1(b),
in terms of manipulating the parameter vector $\vec{R}$.
For a spin-1/2 particle in magnetic field,
the vector in the parameter space
just corresponds to the Bloch vector of the spin state,
i.e., the eigenstate of $\sigma_n=\vec{\sigma}\cdot \vec{n}$,
with $\vec{n}$ the unit vector of $\vec{R}$.
This would result in the well-known Berry phase characterized by
the solid-angle spanned to the closed trajectory in the parameter space.
For the case of Majorana braiding as shown in Fig.\ 1 (b),
it will be more difficult to understand the similar result,
despite that one can indeed obtain the same Berry phase
by means of certain more sophisticated treatment
\cite{Fre16,Xie21}.

Moreover, in order to handle the nonadiabatic effect
during evolution along the parameter trajectory shown in Fig.\ 1(b),
we would like to describe the state evolution
using the occupation-number-state representation.
In general, the state can be expressed as
\bea
|\Psi(t)\ra = \alpha(t) |E_-(t)\ra + \beta(t) |E_+(t)\ra  \,,
\eea
where $|E_-(t)\ra$ and $|E_+(t)\ra$ are the instantaneous eigenstates
of the two-state system under study.
In the adiabatic case, we always have $\beta(t)=0$,
since there is no nonadiabatic transition occurred.
The system will always stay in the instantaneous eigenstate
connected with the initial state, say, $|0_{23}0_d\ra$ assumed here.
We thus have $|\alpha(t)|=\alpha(0)=1$.
At the end of braiding,
$\alpha(T)= e^{i\varphi} \alpha(0)$.
The most important point is that the phase factor $\varphi$
does not contain only the expected dynamical phase
$\varphi_{\rm dyn}= -\frac{1}{\hbar}\int^{T}_{0}dt E_{-}(t)$,
but also contain a geometric phase $\varphi_g$, i.e,
the Berry phase determined by the solid angle as shown in Fig.\ 1(b).
Therefore, in total, we have
$\varphi = \varphi_{\rm dyn} + \varphi_g$.
In the presence of nonadiabatic transition,
in addition to $\beta(t)\neq 0$,
the remained part from the phase $\varphi(T)$
after subtracting the dynamic phase $\varphi_{\rm dyn}(T)$,
will also deviate from the geometric phase $\varphi_g$
determined by the solid angle spanned in the parameter space.
We will show that the both errors
will affect the fidelity of the braiding operation.

In Fig.\ 2 we show the numerical results simulated based on consideration outlined above.
The scheme of four-step uniform modulation of the coupling strengths
can be summarized as $|d\lambda_{1,2}/dt|=\frac{\lambda_c}{T/4}$,
which are involved sequentially in the following modulations:
(i) with $|\lambda_1|$ increased from zero to $\lambda_c$;
(ii) with $|\lambda_2|$ increased from zero to $\lambda_c$;
(iii) with $|\lambda_1|$ decreased from $\lambda_c$ to zero;
and (iv) with $|\lambda_2|$ decreased from $\lambda_c$ to zero.
In Fig.\ 2(a) and (b) we show the results
of the nonadiabatic transition probability $|\beta|^2$
and the geometric phase error (deviated from the Berry phase $\pi/4$),
{\it versus} the braiding time $T$.
In general, as expected, the nonadiabatic transition
is less prominent with increase of $T$.
The behavior of oscillations observed in Fig.\ 2(a) is owing to
the Landau-Zener-St\"uckelberg (LZS) interference \cite{Lan32,Zen32,Nori10}
between the (relatively strong) nonadiabatic transitions
near $t=0$ and $t=T$, where the energy gap is small.
In Fig.\ 2(b), we find that the geometric phase
extracted from the dynamic (nonadiabatic) evolution
deviates also from the result of the adiabatic case,
i.e., the geometric phase $\varphi_g=\Omega_c/2$,
while the solid angle takes the value of
$\Omega_c = \arcsin [4\lambda_c^2/ (\varepsilon_d^2+4\lambda_c^2)]$,
numerically which corresponds to
the asymptotic result in the long $T$ limit in Fig.\ 2(b).

The two errors shown in Fig.\ 2(a) and (b)
will affect the fidelity of the logic gate
associated with the braiding operation.
In ideal case with no errors,
the desired state of the 4-MZMs qubit after braiding is $|\psi_g\ra$,
as shown in \Eq{psi-g}.
The fidelity compared with this state
is given by ${\cal F}={\rm Tr}_M(|\psi_g\ra \la\psi_g|\rho_M)$,
where $\rho_M={\rm Tr}_{D}[|\Psi(T)\ra \la\Psi(T)|]$
is the reduced density matrix
after tracing out the QD degree of freedom from the total state $|\Psi(T)\ra$,
thus the remained trace ${\rm Tr}_M(\cdots)$ is over the Majorana qubit states.
In Fig.\ 2(c), we show the numerical result of the fidelity.
We find that the fidelity is largely affected
by the nonadiabatic transition probability $|\beta|^2$ shown in Fig.\ 2(a).

We know that the relatively strong nonadiabatic transitions
largely take place near $t=0$ and $t=T$,
where the energy gap is small as shown in Fig.\ 1(c).
Qualitatively speaking, in order to avoid strong nonadiabatic transition,
one should modulate the parameter change more slowly
for smaller energy difference $\Delta E=E_+-E_-$.
Therefore, let us consider a nonuniform change of the parameters
in the four-step modulation scheme according to
$|d\lambda_{1,2}/dt|=\eta (E_+-E_-)^2/\hbar$,
where the dimensionless parameter $\eta$
is introduced to control the parameter modulation speed.
If $\eta<<1$, the well-know adiabatic condition is satisfied.
On the contrary, if $\eta >1$,
remarkable nonadiabatic transition will take place.
For this nonuniform scheme of parameter modulation,
it is possible to carry out an analytic expression
for the total time of braiding.
The individual times for each of the four steps can be obtained as
\bea
T_1&=&T_4=\frac{1}{2\gamma \varepsilon_d}\arctan{\frac{2\lambda_c}{\varepsilon_d}}  \,, \nl
T_2&=&T_3=\frac{1}{2\gamma \sqrt{\varepsilon_d^2+4\lambda_c^2}}\arctan{\frac{2\lambda_c}
{\sqrt{\varepsilon_d^2+4\lambda_c^2}}}  \,.
\eea
The total braiding time is their sum, $T=T_1+T_2+T_3+T_4$.
In Fig.\ 2(a), (b) and (c) we plot also the results
of the nonuniform modulation scheme,
in close comparison with the results of the uniform scheme.
We find that, remarkably, the nonadiabatic transition
can be largely suppressed,
with thus an important advantage
of allowing much shorter braiding times.

\begin{figure}
  \centering
  \includegraphics[scale=0.4]{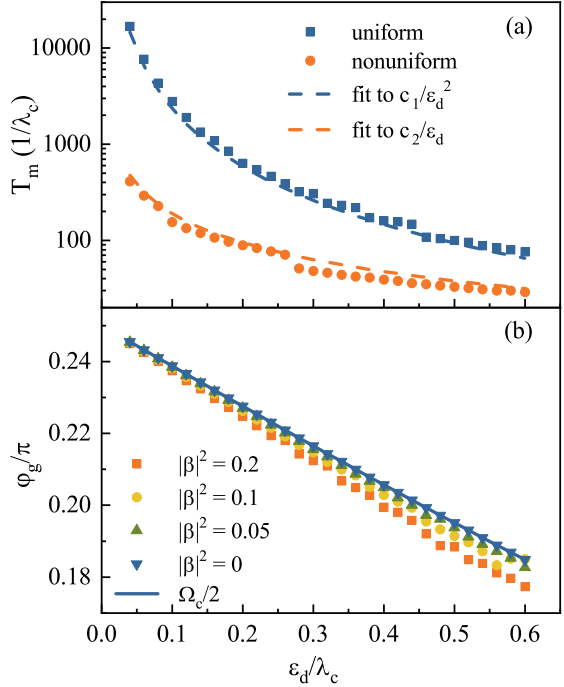}
  \caption{ (a)
Minimum braiding time $T_m$ under the condition $|\beta|^2 \leq 0.1$
{\it versus} the dot energy level $\varepsilon_d$,
for the four-step uniform and nonuniform manipulation schemes.
The numerical results are fitted by using
$c_1/\varepsilon_d^2$ and $c_2/\varepsilon_d$, respectively,
 while $c_1=32\ln10/\pi$ is obtained through the formula of
Landau-Zener transition probability
and $c_2\approx 19$ is obtained through numerical fitting.
(b)
Geometric phases (plotted by symbols) extracted from dynamic evolutions
within the respective braiding time $T_m$ under the specified threshold conditions.
The blue line plots the analytic result of the Berry phase  $\Omega_c/2=\frac{1}{2}\arcsin[4\lambda_c^2/(\varepsilon_d^2+4\lambda_c^2)]$ .  }
\end{figure}

In Fig.\ 3 we show further
the particular effect of the dot energy $\varepsilon_d$.
In Fig.\ 3(a), the dependence of the minimum braiding time $T_m$
on the dot energy $\varepsilon_d$ is shown
for both the uniform and nonuniform modulation schemes.
The minimum braiding time $T_m$ is determined from the threshold value
of the nonadiabatic transition probability $|\beta(T)|^2<0.1$ if $T>T_m$.
Again, we find that the results of the nonuniform modulation scheme
are much better than the uniform scheme.
This becomes more prominent with decrease of the dot energy.
In Fig.\ 3(b) we show the geometric phase {\it versus} the dot energy $\varepsilon_d$,
where the results (symbols) extracted from the real time dynamic evolution
are plotted against the adiabatic value (blue line)
determined by the solid angle as shown in Fig.\ 1(b).

\begin{figure}
  \centering
  \includegraphics[scale=0.8]{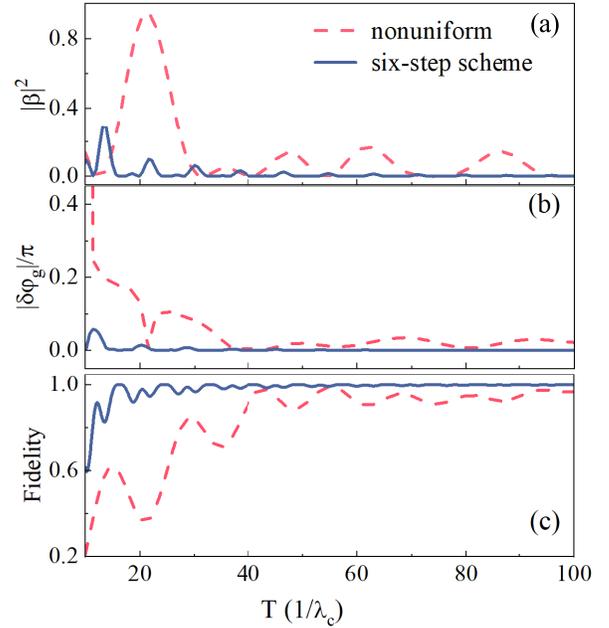}
  \caption{ Nonadiabatic effect of the improved six-step manipulation scheme,
in comparison with the relatively good results of
the four-step nonuniform scheme as shown in Fig.\ 2.
Plotted in (a), (b) and (c) are the braiding time dependence,
respectively, for the transition probability,
the geometric phase error from $\pi/4$,
and the fidelity of the 4-MZMs qubit state rotation.
In the simulations, we assume $\varepsilon_d=0.2\lambda_c$
in the whole process of the four-step manipulation,
and $\varepsilon_d=2\lambda_c$ as the initial and final values of the dot energy
for the six-step scheme.  }
\end{figure}

\vspace{0.1cm}
{\flushleft\it Six-step scheme}. ---
In order to avoid strong nonadiabatic transitions
near the beginning and end of the braiding operation,
we need a relatively large dot energy, e.g., $\varepsilon_d=2\lambda_c$.
In order to achieve at the same time
a complete braiding between $\gamma_2$ and $\gamma_3$,
which requires to realize an exact solid angle of $\Omega_c=\pi/2$
as schematically shown in Fig.\ 1(b),
we further propose the following six-step modulation protocol:
(i) with $|\lambda_1|$ increased from zero to $\lambda_c$;
(ii) with $\varepsilon_d$ decreased from $2\lambda_c$ to zero;
(iii) with $|\lambda_2|$ increased from zero to $\lambda_c$;
(iv) with $|\lambda_1|$ decreased from $\lambda_c$ to zero;
(v) with $|\lambda_2|$ decreased from $\lambda_c$ to zero;
and (vi) with $\varepsilon_d$ restored the initial value from zero to $2\lambda_c$.
For each step, the modulation speed can be uniform,
say, to modulate the energy with a constant rate.

In Fig.\ 4 we compare the results of this six-step scheme
with the four-step protocol results shown in Fig.\ 2.
As a more challenging comparison,
we compare here only with the nonuniform four-step scheme,
which has been demonstrated with better effect of
suppressing the nonadiabatic transition.
In Fig.\ 4(a), we find that the nonadiabatic transition can be
largely suppressed even on short timescale of braiding,
while in Fig.\ 4(b) and (c)
we find the precise geometric phase $\varphi_g=\pi/4$
and good fidelity with the desired state $|\psi_g\ra$ accomplished,
which indicate a complete exchange of the MZMs $\gamma_2$ and $\gamma_3$.
Importantly, the uniform six-step modulation should be more feasible in practice
than the adiabatic-condition-guided nonuniform four-step scheme analyzed above.
Here, only an additional modulation of the dot energy $\varepsilon_d$ is added.
This can be done similarly as modulating the coupling strengths via electric gate control.

\begin{figure}
  \centering
  \includegraphics[scale=0.8]{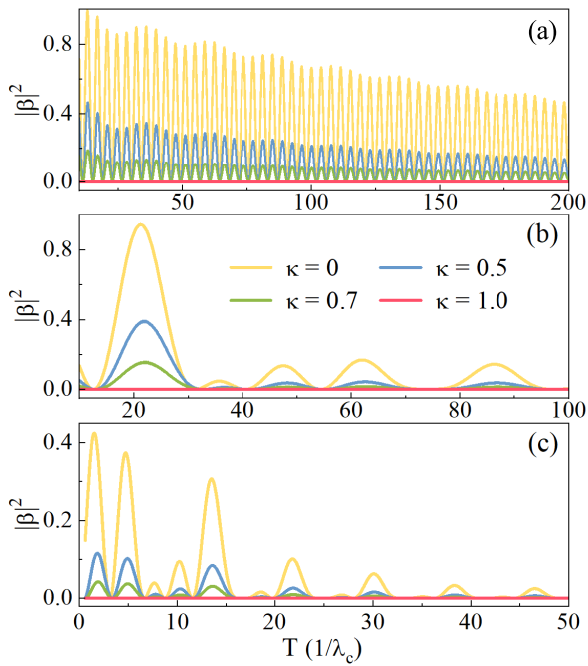}
  \caption{ Effect of counterdiabatic driving,
by varying its action strength $\kappa$,
based on the shortcuts-to-adiabaticity approach.
The nonadiabatic transition probabilities
shown in (a), (b) and (c) correspond to
the four-step uniform, nonuniform
and the six-step manipulation schemes, respectively.
The quantum dot energy level $\varepsilon_d$
is assumed the same as explained in the caption of Fig.\ 4.    }
\end{figure}

\vspace{0.1cm}
{\flushleft\it Shortcuts-to-adiabaticity approach}. ---
Finally, we notice that a popular method of suppressing the nonadiabatic transition
is the so-called shortcuts-to-adiabaticity (STA) approach \cite{Rice03,Ber09,Opp15,Rah18}.
The basic idea of STA is to add auxiliary counterdiabatic driving
so that the dynamics can follow the adiabatic evolution.
For the two-state system under present study,
as shown in Fig.\ 1(b) and (c),
the counterdiabatic driving Hamiltonian can be obtained as
$H'(t) = i\hbar (\zeta_1\gamma_B\gamma_3+\zeta_2\gamma_B\gamma_2
+\zeta_3\gamma_2\gamma_3) / \Delta^2$,
where
$\Delta^2=\varepsilon_d^2+4|\lambda_1|^2+4|\lambda_2|^2$,
$\zeta_j=(-1)^j (|\dot{\lambda}_j|\varepsilon_d - \dot{\varepsilon}_d|\lambda_j|)$
for $\zeta_1$ and $\zeta_2$,
and $\zeta_3=2(|\lambda_1 \dot{\lambda}_2|-|\dot{\lambda}_1 \lambda_2|) $.
Then, the total Hamiltonian to affect the system evolution is
$H(t)=H_M(t)+\kappa H'(t)$.
Here we introduce an overall strength parameter $\kappa$
for the counterdiabatic driving,
in order to demonstrate the effect of the STA approach.
In Fig.\ 5 we display the effect of the STA counterdiabatic driving
on the nonadiabatic transition probabilities
for the uniform four-step, nonuniform four-step
and uniform six-step schemes in (a), (b) and (c), respectively,
on different timescales based on the previous exiting results.
For each case, we consider the counterdiabatic driving strengths of $\kappa=0$, 0.5, 0.7 and 1.
As expected, nonadiabatic transition can be suppressed for every case
if we perform the full control with $\kappa=1$.
The results shown in Fig.\ 5 are indeed satisfactory.
However, one can easily find that
the counterdiabatic driving Hamiltonian
is very hard to be realized in practice.
Here, in addition to switching on coupling between $\gamma_B$ and $(\gamma_2, \gamma_3)$,
we also need to establish coupling between $\gamma_2$ and $\gamma_3$.
All these couplings are beyond the original manipulation couplings as shown in Fig.\ 1(b).
Moreover, the most serious difficulty is that
precise realization of these time-dependent coupling strengths
is seemingly impossible in practice.
Deviations from these mathematically designed coupling strengths
will result in a failure to the success
as roughly shown in Fig.\ 5.

\vspace{0.1cm}
{\flushleft\it Summary and discussion}. ---
By simulating the real time dynamics of braiding a pair of MZMs
through a quantum dot, we revealed the strong nonadiabatic effect
when the dot energy level approaches to zero
as required for a full exchange between the MZMs.
We simulated and compared the results of different schemes.
Rather than nonuniformly manipulating the system according to
adiabatic condition and shortcuts-to-adiabaticity strategy,
we proposed a simpler and more feasible scheme of six-step manipulations
to suppress the nonadiabatic transition,
which allows as well for a full exchange between the Majorana modes.

The Majorana braiding protocol analyzed in this work
(with the help of a quantum dot)
is quite compatible with the nowadays
hybrid semiconductor-superconductor experimental platforms.
It is also the minimal 1D realization of braiding a pair of MZMs.
We may highlight that the scheme of the six-step manipulations
has obvious advantages over the nonuniform four-step scheme
and the shortcuts-to-adiabaticity approach,
since it should be very difficult in practice to implement
the time-dependent precise controls required by the latter schemes.
The nonadiabatic transitions can be inferred via the dot electron occupation,
which can be further detected by a nearby charge sensitive detector
such as quantum-point-contact device.
Moreover, the braiding resultant state, given by \Eq{psi-g}, and deviations from it,
can be verified by moving (via gate-voltage-controls)
the Majorana modes $\gamma_2$ and $\gamma_3$
to the other ends of the wires, to fuse with $\gamma_1$ and $\gamma_4$.
Then, one can measure the occupations $n_{12}$ and $n_{34}$
of the fused regular fermions, by coupling them to
individual quantum dots and quantum-point-contact detectors,
following the proposal in Ref.\ \cite{NC22} to initialize the Majorana pairs
for demonstrating the nontrivial fusion of Majorana fermions.
We believe that along this line,
based on the platform of quantum dot coupled Majorana wires,
demonstrating the non-Abelian statistics of MZMs
should be an attractive research direction.

\vspace{0.2cm}
{\flushleft\it Acknowledgements.}---
This work was supported by
the NNSF of China (Nos.\ 11974011 \& 11904261).


\end{document}